\documentclass[preprint,preprintnumbers, prd, floatfix,  superscriptaddress,nofootinbib] {revtex4-1}
\usepackage{graphicx}
\usepackage{epsfig}
\usepackage{bm}
\usepackage{amssymb}
\usepackage{float}
\usepackage{amsmath}
\usepackage{dcolumn}
\usepackage{cancel}
\usepackage{color}
\usepackage{epstopdf}
\usepackage{nccbbb}

\begin{document}

\title {Relieve the $H_0$ tension with a new coupled generalized three-form dark energy model}

\author{Yan-Hong Yao}
\email{yhy@mail.nankai.edu.cn}
\author{Xin-He Meng}
\email{xhm@nankai.edu.cn}

\affiliation{Department of Physics, Nankai University, Tianjin 300071, China}

\begin{abstract}
In this work we propose a new coupled generalized three-form dark energy model, in which dark energy are represented by a three-form field and other components are represented by ideal fluids. We first perform a dynamical analysis on the new model and obtain four fixed points, including a saddle point representing a radiation dominated Universe, a saddle point representing a matter dominated Universe, and two attractors representing two dark energy dominated Universes. We then use the observational data, including cosmic microwave background (CMB) data, baryon acoustic oscillations (BAO) data, and Type Ia supernovae (SN Ia) data to constrain the model parameters of the coupled generalized three-form dark energy model. For comparison, we also consider the coupled three-form dark energy model, generalized three-form dark energy model, and $\Lambda$CDM model, we find that the coupled generalized three-form dark energy model is the only one model that can reduce the $H_0$ tension to a more acceptable level, with $H_0=70.1_{-1.5}^{+1.4}$ km/s/Mpc, which is consistent with R19 at $2.0\sigma$ confidence level. We also investigate the best-fit dynamical behavior of the coupled generalized three-form dark energy model, and show that our model is equivalent to a quintom dark energy model, in which dark energy, at early epoch, behaves like some form of early dark energy with a small positive equation of state.

\textbf{}
\end{abstract}

\maketitle

\section{Introduction}
\label{intro}
Despite the canonical $\Lambda$CDM has proven to provide an excellent fit to a wide range of cosmological data\cite{riess1998observational,perlmutter1999measurements,dunkley2011atacama,hinshaw2013nine,ade2014planck,story2015measurement,ade2016planck,alam2017clustering,troxel2018dark,aghanim2018planck}, there are some tensions among the values of cosmological parameters inferred from different datasets. The most striking one is between the value of the Hubble constant $H_0$ inferred from Planck Cosmic Microwave Background (CMB) data and direct local distance ladder measurements. The former has been measured by the Planck satellite, with the Planck collaboration reporting $H_0=67.4\pm0.5$ km/s/Mpc \cite{aghanim2018planck}, which is at 4.4$\sigma$ difference to value measured by the Hubble Space Telescope (HST), $H_0=74\pm1.4$ km/s/Mpc(measurement denoted as R19 hereafter) \cite{riess2019large}.

Since it is unlikely that the Planck observation and/or the local distance ladder measurements suffer from considerable  unaccounted systematic errors \cite{spergel2015planck,addison2016quantifying,aghanim2017planck,cardona2017determining,follin2018insensitivity}, increasing attention is focusing on the possibility that $H_0$ tension might be the first sign for physics beyond $\Lambda$CDM. The most economical modifications to the concordance model include replacing the cosmological constant $\Lambda$ by a dark energy component with equation of state $w<-1$ \cite{Li2013Planck,huang2016how} (i.e. phantom dark energy) and introducing some form of dark radiation \cite{battye2014evidence,Zhang2014Neutrinos,zhang2015sterile,feng2018searching,zhao2018measuring,choudhury2019constraining} (so as to raise the extra relativistic degrees of freedom $N_{eff}$ beyond 3.046). Although these two modifications can help with the $H_0$ tension, it is worth to mention that phantom dark energy and extra relativistic species are disfavored from both baryonic acoustic oscillations (BAO) and Type Ia supernovae (SN Ia) data and from a model comparison point of view \cite{vagnozzi2020new}. In recent years, lots of other scenarios attempting to solve the $H_0$ tension have been examined, including (but ont limited to) an exotic early dark energy that acts as a cosmological constant before a critical redshift $z_c$ but whose density then dilutes faster than radiation \cite{agrawal2019rock,poulin2019early,smith2020oscillating,lin2019acoustic}, interactions between dark matter and dark energy \cite{di2017can,yang2018interacting,di2020nonminimal,di2020interacting,cheng2020testing,lucca2020tensions,gomez2020update}, a vacuum phase transition \cite{2018Vacuum,di2020h0}, modifications to gravity \cite{khosravi2019h,nunes2018structure,cai2019model}.

In Ref.\cite{yao2020new}, we put forward a new coupled three-form dark energy model, and find that its prediction of the Hubble constant $H_0$ inferred from CMB, BAO, SN Ia, and cosmic chronometers (CC) datasets is still in strong tension with the latest local measured value of $H_0$. Since an uncoupled three-form field with vanishing potential is equivalent to a cosmological constant with equation of state as $-1$, inspired by the coupled phantom dark energy model that is successful to relieve the Hubble tension, we generalize the coupled three-form dark energy model proposed in \cite{yao2020new} to a coupled noncanonical three-form dark energy model to address the $H_0$ tension. Noncanonical three-form field, also called as generalized three-form field, has already introduced in\cite{PhysRevD.96.023516,Wongjun2017Generalized,yao2020coupled} to compare with k-essence.

The rest of this paper is organized as follows. In section \ref{sec:1}, we present a new coupled generalized three-form dark energy model in which dark energy are represented by a three-form field and other components are represented by ideal fluids.
In section \ref{sec:2}, we use a dynamical system approach to study the coupled generalized three-form dark energy model through the stability of their fixed points. In section \ref{sec:3}, we constrain the model with the data from CMB, BAO, and SN Ia observations and assess its ability to address the $H_0$ tension. In the last section, we make a brief conclusion with this paper.
\section{A coupled generalized three-form dark energy model}
\label{sec:1}
In this section we generalize the coupled three-form dark energy model proposed in \cite{yao2020new} to a coupled noncanonical three-form dark energy model. We restrict the coupling between two dark sectors to be the conformal form \cite{koivisto2013coupled,yao2018a,yao2020new,yao2020coupled}, a case that has been thoroughly studied in the context of scalar fields. The total Lagrangian is written as
\begin{equation}
\mathcal{L}=\frac{R}{2\kappa^2}+\mathcal{L}_{(d)}+\mathcal{L}_{(b)}+\mathcal{L}_{(\gamma)}+\mathcal{L}_{(\nu)},
\end{equation}
where $R$ denotes the Ricci scalar and $\kappa=\sqrt{8\pi G}$ is the inverse of the reduced Planck mass.
$d$, $b$, $\gamma$, and $\nu$ denote dark sectors, baryon, photon, and neutrino, respectively. For ideal fluids, it is complicated to derive the equations of motion from a variational principle since the constraint equations satisfied by the fluid variables is also needed to be considered. Several variational formulations have been proposed to solve this problem. In this work we consider the variational formulations discussed in \cite{ray1972lagrangian}, then each Lagrangian can be expressed as
\begin{eqnarray}
\mathcal{L}_{(d)} &=& -\frac{1}{48}F^{2}N(A^{2})-I(A^{2})\tilde{\rho}_{(c)}+\lambda_{1}(g_{\mu\nu}u_{(c)}^{\mu}u_{(c)}^{\nu}+1)+\lambda_{2}\nabla_{\alpha}(\tilde{\rho}_{(c)}u_{(c)}^{\alpha}), \\
\mathcal{L}_{(b)} &=&-\rho_{(b)}+\lambda_{3}(g_{\mu\nu}u_{(b)}^{\mu}u_{(b)}^{\nu}+1)+\lambda_{4}\nabla_{\alpha}(\rho_{(b)}u_{(b)}^{\alpha}), \\
\mathcal{L}_{(\gamma)} &=&
-\tilde{\rho}_{(\gamma)}(1+\epsilon_{(\gamma)}(\tilde{\rho}_{(\gamma)}))+\lambda_{5}(g_{\mu\nu}u_{(\gamma)}^{\mu}u_{(\gamma)}^{\nu}+1)+\lambda_{6}\nabla_{\alpha}(\tilde{\rho}_{(\gamma)}u_{(\gamma)}^{\alpha}), \\
\mathcal{L}_{(\nu)} &=& -\tilde{\rho}_{(\nu)}(1+\epsilon_{(\nu)}(\tilde{\rho}_{(\nu)}))+\lambda_{7}(g_{\mu\nu}u_{(\nu)}^{\mu}u_{(\nu)}^{\nu}+1)+\lambda_{8}\nabla_{\alpha}(\tilde{\rho}_{(\nu)}u_{(\nu)}^{\alpha}),
\end{eqnarray}
$A$ and $F=dA$ represent the three-form field and the field strength tensor, $N(A^2)$ and $I(A^2)$ contain total coupling information, including the information of self-interaction of the three-form field and interaction between two dark sectors, respectively. $\lambda_1-\lambda_8$ are the multipliers. $\tilde{\rho}$ denotes the rest density, and $\epsilon$ denotes the rest, specific internal energy, it is connected with pressure $p$ through the following relation,
\begin{eqnarray}
  \frac{d\epsilon}{d\tilde{\rho}}&=& \frac{p}{\tilde{\rho}^{2}}.
\end{eqnarray}

Providing with the total Lagrangian, one can write down the following total action.
\begin{equation}\label{}
 S=\int\mathcal{L}\sqrt{-g}d^{4}x.
\end{equation}
Varying the action with respect to $g^{\mu\nu},\tilde{\rho}_{(c)},u_{(c)}^{\alpha},\rho_{(b)},u_{(b)}^{\alpha},\tilde{\rho}_{(\gamma)},u_{(\gamma)}^{\alpha},\tilde{\rho}_{(\nu)},u_{(\nu)}^{\alpha}$ , we have
\begin{eqnarray}
R_{\mu\nu}-\frac{1}{2}g_{\mu\nu}R &=& \kappa^{2}T_{\mu\nu}, \\
(\partial_{\alpha}\lambda_{2})u_{(c)}^{\alpha} &=& -I, \\
\lambda_{1} &=& \frac{1}{2}I\tilde{\rho}_{(c)}, \\
(\partial_{\alpha}\lambda_{4})u_{(b)}^{\alpha} &=& -1, \\
\lambda_{3} &=& \frac{1}{2}\rho_{(b)}, \\
(\partial_{\alpha}\lambda_{6})u_{(\gamma)}^{\alpha} &=& -(1+\epsilon_{(\gamma)}(\tilde{\rho}_{(\gamma)})),\\
\lambda_{5} &=& \frac{1}{2}(1+\epsilon_{(\gamma)}(\tilde{\rho}_{(\gamma)}))\tilde{\rho}_{(\gamma)}, \\
(\partial_{\alpha}\lambda_{8})u_{(\nu)}^{\alpha} &=& -(1+\epsilon_{(\nu)}(\tilde{\rho}_{(\nu)})), \\
\lambda_{7} &=& \frac{1}{2}(1+\epsilon_{(\nu)}(\tilde{\rho}_{(\nu)}))\tilde{\rho}_{(\nu)},
\end{eqnarray}
thanks to the equations $(9)-(16)$, the total energy-momentum tensor for all components can be written as
\begin{equation}\label{}
\begin{split}
 T_{\mu\nu}=&\frac{1}{6}NF_{\mu\alpha\beta\gamma}F_{\nu}^{\alpha\beta\gamma}+6(\frac{1}{48}\frac{dN}{dA^{2}}F^{2}+\frac{dlnI}{dA^{2}}\rho_{(c)})A_{\mu}^{\alpha\beta}A_{\nu\alpha\beta}-g_{\mu\nu}\frac{1}{48}F^{2}N\\
 &+\rho_{(c)}u_{(c)\mu}u_{(c)\nu}+\rho_{(b)}u_{(b)\mu}u_{(b)\nu}+(\rho_{(\gamma)}+p_{(\gamma)})u_{(\gamma)\mu}u_{(\gamma)\nu}+p_{(\gamma)}g_{\mu\nu}+
 (\rho_{(\nu)}+p_{(\nu)})u_{(\nu)\mu}u_{(\nu)\nu}+p_{(\nu)}g_{\mu\nu},
\end{split}
\end{equation}
where $\rho_{(c)}=I(A^{2})\tilde{\rho}_{(c)}$, $\rho_{(\gamma)}=\tilde{\rho}_{(\gamma)}(1+\epsilon_{(\gamma)}(\tilde{\rho}_{(\gamma)}))$, and $\rho_{(\nu)}=\tilde{\rho}_{(\nu)}(1+\epsilon_{(\nu)}(\tilde{\rho}_{(\nu)}))$.
Since the Universe is assumed as homogeneous and isotropic, we have $u_{\mu}=u_{(c)\mu}=u_{(b)\mu}=u_{(\gamma)\mu}=u_{(\nu)\mu}$. Furthermore, we denote $\rho_{(r)}=\rho_{(\gamma)}+\rho_{(\nu)}$ and $p_{(r)}=p_{(\gamma)}+p_{(\nu)}$. Therefore, (17) becomes to
\begin{equation}\label{}
\begin{split}
 T_{\mu\nu}=&\frac{1}{6}NF_{\mu\alpha\beta\gamma}F_{\nu}^{\alpha\beta\gamma}+6(\frac{1}{48}\frac{dN}{dA^{2}}F^{2}+\frac{dlnI}{dA^{2}}\rho_{(c)})A_{\mu}^{\alpha\beta}A_{\nu\alpha\beta}-g_{\mu\nu}\frac{1}{48}F^{2}N\\
 &+\rho_{(c)}u_{\mu}u_{\nu}+\rho_{(b)}u_{\mu}u_{\nu}+(\rho_{(r)}+p_{(r)})u_{\mu}u_{\nu}+p_{(r)}g_{\mu\nu}.
\end{split}
\end{equation}
The variation of the total action with respect to the three-form field leads to the following equations of motion
\begin{equation}\label{}
  (\nabla_{\alpha}N)F^{\alpha\mu\nu\rho}+N\nabla_{\alpha}F^{\alpha\mu\nu\rho} = 12(\frac{1}{48}\frac{dN}{dA^{2}}F^{2}+\frac{dlnI}{dA^{2}}\rho_{(c)})A^{\mu\nu\rho}
\end{equation}
Noting that $N(A^2)$ is independent from $F^2$, so all of the second derivatives terms is contained in the second term at the left side of the equation (19), therefore the noncanonical field theory proposed in this work is satisfied with Hyperbolicity condition in a similar way with the canonical field theory.

Using the equations of motion for the three-form field and the vanishing of the divergence of the stress energy tensor for two dark sectors we have the equation of motion for dark matter:
\begin{equation}\label{}
  \nabla_{\mu}(\rho_{(c)}u^{\mu}u_{\nu})=-2\frac{dlnI}{dA^{2}}\rho_{(c)}A^{\alpha\beta\gamma}\nabla_{\nu}A_{\alpha\beta\gamma}.
\end{equation}

The Universe is assumed as homogeneous, isotropic, and spatially flat in this work, as a result, it is described by the Friedmann-Robertson-Walker (FRW) metric,
\begin{equation}\label{}
  ds^{2}=-dt^{2}+a(t)^{2}d\vec{x}^{2},
\end{equation}
here $a(t)$ stands for the scale factor.
To be compatible with FRW symmetries, the three-form field is chosen as the time-like component of the dual vector field\cite{Koivisto2009Inflation2}, i.e.
\begin{equation}
  A_{i j k}=X(t)a(t)^{3}\varepsilon_{ijk}.
\end{equation}

To specify a coupled generalized three-form dark energy model, we assume $N(A^2)$ and $I(A^2)$ to be
\begin{eqnarray}\label{}
  N &=&(\alpha^2+\frac{\kappa^{2}}{6}A^{2})^{\frac{\alpha}{2}}=(\alpha^2+\kappa^{2}X^{2})^{\frac{\alpha}{2}} \hspace{1cm} \\
  I &=&(1+\frac{\kappa^{2}}{6}A^{2})^{\frac{\beta}{2}}=(1+\kappa^{2}X^{2})^{\frac{\beta}{2}}
\end{eqnarray}
where $\alpha$ and $\beta$ are two coupling constant.

Now we have the Friedmann equations:
\begin{eqnarray}
  H^{2} &=& \frac{\kappa^{2}}{3}\rho, \\
  \dot{H} &=&-\frac{\kappa^{2}}{2}(\rho+p),
\end{eqnarray}
where
\begin{eqnarray}
  \rho &=&- T_{0}^{0}=-g^{00}T_{00}=\frac{1}{2}(\alpha^2+\kappa^{2}X^{2})^{\frac{\alpha}{2}}(\dot{X}+3HX)^{2}+\rho_{(c)}+\rho_{(b)}+\rho_{(r)}, \\
  p &=&\frac{1}{3}T_{i}^{i}=\frac{1}{3}g^{ii}T_{ii}=-\frac{1}{2}(\alpha^2+\kappa^{2}X^{2})^{\frac{\alpha}{2}}(1+\alpha \frac{\kappa^{2}X^{2}}{\alpha^2+\kappa^{2}X^{2}} )(\dot{X}+3HX)^{2}+\beta\frac{\kappa^{2}X^{2}}{1+\kappa^{2}X^{2}} \rho_{(c)}+\frac{1}{3}\rho_{(r)}.\hspace{1cm}
\end{eqnarray}

In the FRW space-time, there is only one independent equation of motion of the three-form field, i.e.
\begin{equation}\label{}
   (\alpha^2+\kappa^{2}X^{2})^{\frac{\alpha}{2}}(\ddot{X}+3\dot{H}X+3H\dot{X})+\frac{\alpha \kappa^2 X (\alpha^2+\kappa^{2}X^{2})^{\frac{\alpha}{2}-1}}{2}(\dot{X}-3HX)(\dot{X}+3HX)+ \beta\frac{\kappa^{2}X}{1+\kappa^{2}X^{2}} \rho_{(c)}=0,
\end{equation}

in addition, the energy conservation equations of two dark sectors is written as
\begin{eqnarray}
  \dot{\rho}_{(c)}+3H\rho_{(c)} &=& \delta H\rho_{(c)},\\
   \dot{\rho}_{(X)}+3H(\rho_{(X)}+p_{(X)})& = &-\delta H\rho_{(c)},
\end{eqnarray}
where
\begin{eqnarray}
 \delta&=&\beta\frac{\kappa^{2}XX^{\prime}}{1+\kappa^{2}X^{2}}, \\
\rho_{(X)} &=&\frac{1}{2}(\alpha^2+\kappa^{2}X^{2})^{\frac{\alpha}{2}}(\dot{X}+3HX)^{2},\\
p_{(X)} &=&-\frac{1}{2}(\alpha^2+\kappa^{2}X^{2})^{\frac{\alpha}{2}}(1+\alpha \frac{\kappa^{2}X^{2}}{\alpha^2+\kappa^{2}X^{2}} )(\dot{X}+3HX)^{2}+\beta\frac{\kappa^{2}X^{2}}{1+\kappa^{2}X^{2}} \rho_{(c)}.
\end{eqnarray}

\section{The autonomous system of evolution equations }
\label{sec:2}
In order to study the coupled generalized three-form dark energy model in a dynamical system approach, it is convenient to introduce the following dimensionless variable \cite{Koivisto2009Inflation2}
\begin{equation}\label{}
  x_{1}=\kappa X, \hspace{1cm} x_{2}=\frac{\kappa}{\sqrt{6}}(X^{\prime}+3X), \hspace{1cm} x_{3}=\frac{\kappa \sqrt{\rho}_{(b)}}{\sqrt{3}H}, \hspace{1cm} x_{4}=\frac{\kappa \sqrt{\rho_{(r)}}}{\sqrt{3}H},
\end{equation}
with these dimensionless variable, we can write down the following autonomous system of evolution equations.
\begin{equation}
  x_{1}^{\prime}=\sqrt{6}x_{2}-3x_{1}
\end{equation}
\begin{equation}
  \begin{split}
  x_{2}^{\prime} & =(\frac{3}{2}(1+\frac{\beta x_{1}^{2}}{1+x_{1}^{2}})(1-(\alpha^2+x_{1}^2)^{\frac{\alpha}{2}}x_{2}^2-x_{3}^2-x_{4}^2)-\frac{3}{2}\alpha x_1^2 (\alpha^2+x_{1}^2)^{\frac{\alpha}{2}-1}x_{2}^2+\frac{3}{2}x_{3}^2+2x_{4}^2)x_{2}\hspace{1cm} \\
    & -\frac{\sqrt{6}\alpha x_1}{2 (\alpha^2+x_{1}^2)}x_{2}^2+3\alpha \frac{x_1^2}{\alpha^2+x_1^2}x_{2}
 -\frac{\sqrt{6}}{2} \frac{\beta x_1}{(\alpha^2+x_{1}^2)^{\frac{\alpha}{2}}(1+x_{1}^2)}(1-(\alpha^2+x_{1}^2)^{\frac{\alpha}{2}}x_{2}^2-x_{3}^2-x_{4}^2),
\end{split}
\end{equation}
\begin{equation}
  x_{3}^{\prime}= -\frac{3}{2}x_{3}+(\frac{3}{2}(1+\frac{\beta x_{1}^{2}}{1+x_{1}^{2}})(1-(\alpha^2+x_{1}^2)^{\frac{\alpha}{2}}x_{2}^2-x_{3}^2-x_{4}^2)-\frac{3}{2}\alpha x_1^2 (\alpha^2+x_{1}^2)^{\frac{\alpha}{2}-1}x_{2}^2+\frac{3}{2}x_{3}^2+2x_{4}^2)x_{3},\hspace{1cm}
\end{equation}
\begin{equation}
 x_{4}^{\prime}= -2x_{4}+(\frac{3}{2}(1+\frac{\beta x_{1}^{2}}{1+x_{1}^{2}})(1-(\alpha^2+x_{1}^2)^{\frac{\alpha}{2}}x_{2}^2-x_{3}^2-x_{4}^2)-\frac{3}{2}\alpha x_1^2 (\alpha^2+x_{1}^2)^{\frac{\alpha}{2}-1}x_{2}^2+\frac{3}{2}x_{3}^2+2x_{4}^2)x_{4}.\hspace{1cm}
\end{equation}
The prime stands for the derivative with respect to e-folding time $N$=ln$a(t)$ here and in the following.

\begin{table}[t]
\begin{center}
  \begin{tabular}{|c|c|c|c|c|c|c|c|}
\hline
&  & &&&&&\\[-.5em]
&$(x_1,x_2,x_3,x_4)$& $\Omega_{(X)}$&$\Omega_{(c)}$&$\Omega_{(b)}$&$\Omega_{(r)}$&$w_{(X)}$&$w_{eff}$
\\[.5em]
\hline
& & & & &&& \\[-.5em]
(a)&$(0,0,0,1)$&$0$&$0$&0&$1$&$-$&$\frac{1}{3}$
\\[.5em]
\hline
& & & & & && \\[-.5em]
(b)&$(0,0,x_3,0)$&$0$&$1-x_3^{2}$&$x_3^{2}$&0&$-$&$0$
\\[.5em]
\hline
& & & & &&& \\[-.5em]
(c)&$(\xi,\sqrt{\frac{3}{2}}\xi,0,0)((\alpha^2+\xi^2)^{\alpha/2}\xi^2=\frac{2}{3})$&$1$&$0$&$0$&$0$&$-1-\alpha\frac{\xi^2}{\alpha^2+\xi^2}$&$-1-\alpha\frac{\xi^2}{\alpha^2+\xi^2}$
\\[.5em]
\hline
& & & & &&& \\[-.5em]
(d)&$(-\xi,-\sqrt{\frac{3}{2}}\xi,0,0)((\alpha^2+\xi^2)^{\alpha/2}\xi^2=\frac{2}{3})$&$1$&$0$&$0$&$0$&$-1-\alpha\frac{\xi^2}{\alpha^2+\xi^2}$&$-1-\alpha\frac{\xi^2}{\alpha^2+\xi^2}$
\\[.5em]
\hline
\end{tabular}
    \caption{Fixed points of the autonomous system of evolution equations.}
\label{fixedpoints}
\end{center}
\end{table}

Fixed points of the autonomous system of evolution equations are presented in Tab.\ref{fixedpoints}. There are four of them, fixed point (a) represents a radiation dominated Universe, it is a saddle point since its eigenvalues are three positive numbers mixed by a negative number, as is showed in the following formula.
\begin{equation}
  \mu_{(a)}=-3,2,1,\frac{1}{2}.
\end{equation}

Fixed point (b) represents a matter dominated Universe, its eigenvalues are
\begin{equation}
  \mu_{(b)}=0,-\frac{1}{2},\frac{-3-\sqrt{3}\sqrt{27-16\beta (\alpha^2)^{-\frac{\alpha}{2}}(1-x_{3}^{2})}}{4},\frac{-3+\sqrt{3}\sqrt{27-16\beta(\alpha^2)^{-\frac{\alpha}{2}}(1-x_{3}^{2})}}{4},
\end{equation}
one of eigenvalues vanishes because fixed point (b) is a line segment consisted of countless points in $x_1=0$, $x_2=0$,
$x_3\in(0,1)$, $x_4=0$ and the eigenvector corresponding to the vanishing eigenvalue is the tangent vector of that line segment. Therefore, the stability of fixed point (b) only depends on other eigenvalues. As a result, fixed point (b) is a saddle point if and only if $\beta<\frac{3(\alpha^2)^{\frac{\alpha}{2}}}{2(1-x_3^2)}$. In the next section, we will show that such condition is satisfied since the fitting results suggest $|\alpha|,|\beta|\ll1$.

Fixed point (c) and fixed point (d) are represented by two dark energy dominated Universes in consideration of $|\alpha|\ll1$, they are always stable since their eigenvalues read
\begin{eqnarray}
  \mu_{(c)} &=& -\frac{3}{2}+\eta_1,-2+\eta_2,-3+\eta_3, -3+\eta_4,\\
  \mu_{(d)} &=&-\frac{3}{2}+\eta_5,-2+\eta_6,-3+\eta_7, -3+\eta_8.
\end{eqnarray}
where $|Re(\eta_i)|\ll1(i=1,2,...,8)$ when $|\alpha|,|\beta|\ll1$.

The phase space is separated in two symmetrical parts because of attractor (c) and attractor (d). In detail, the trajectories in the phase space run toward the attractor (c) if $x_2>\theta(x_1,x_3,x_4)$ at the beginning, otherwise they will run toward the attractor (d), here $\theta(x_1,x_3,x_4)\approx0$.
\section{Confront the model with observations}
\label{sec:3}
In this section we constrain the coupled generalized three-form dark energy model using CMB, BAO, SN Ia datasets based on the following Hubble parameter.
\begin{equation}
 H^2
 =\frac{\Omega_{(r)}(1+z)^{4}+\Omega_{(c)}((1+x_{1}^{2})/(1+x_{10}^{2}))^{\frac{\beta}{2}}(1+z)^{3}+\Omega_{(b)}(1+z)^{3}}{1-(
 \alpha^2+x_{1}^2)^{\frac{\alpha}{2}}x_2^2}.
\end{equation}

\subsection{CMB measurements}
Most of the information contained in the CMB power spectrum can be compressed into two shift parameters. The first, $R$, is defined as
\begin{equation}
  R=\sqrt{\Omega_{(m)}H_0^2}D_{M}(z_{\ast}),
\end{equation}
and the second, $l_{A}$, is given by
\begin{equation}
  l_{A}=\pi\frac{D_{M}(z_{\ast})}{r_{s}(z_{\ast})},
\end{equation}
where $D_{M}=(1+z_{\ast})D_{A}=\int_{0}^{z_{\ast}}\frac{dz}{H}$ and $r_{s}=\int_{0}^{t_{\ast}}\frac{c_s dt}{a(t)}=\frac{1}{\sqrt{3}}\int_{0}^{a_{\ast}}\frac{da}{a^{2}H(a)\sqrt{1+\frac{3\omega_{b}a}{4\omega_{\gamma}}}}$ are the comoving angular distance at decoupling and sound horizon at decoupling, respectively\cite{Efstathiou2010Cosmic}. The former depends on the dominant components after decoupling, while the latter depends on the dominant components before decoupling. $m$ denotes matter, including baryon and dark matter.
The redshift at decoupling $z_{\ast}$ is given by \cite{Hu1996Small}
\begin{eqnarray}
  z_{*}&=&1048(1+0.00124\omega_{(b)}^{-0.738})(1+g_{1}\omega_{(m)}^{g_{2}}),\\
g_{1}&=&\frac{0.0783\omega_{(b)}^{-0.238}}{1+39.5\omega_{(b)}^{0.763}},\\
  g_{2}&=&\frac{0.56}{1+21.1\omega_{(b)}^{1.81}},
\end{eqnarray}

In this work we use the following Planck 2018 compressed likelihood \cite{chen2019distance} with these two shift parameters to perform a likelihood analysis,
\begin{eqnarray}% \nonumber to remove numbering (before each equation)
  \chi_{CMB}^{2}&=&s^{T} C_{CMB}^{-1} s,\\
  s &=& (R-1.7502,l_{A}-301.471,\omega_{(b)}-0.02236),
   \end{eqnarray}
where $C_{ij}=D_{ij}\sigma_i\sigma_j$ is the covariance matrix, $\sigma=(0.0046,0.09,0.00015)$ is the errors, and
$D_{CMB}=  \left(
      \begin{array}{ccc}
        1 & 0.46 &-0.66 \\
       0.46 & 1 & -0.33 \\
       -0.66&-0.33& 1\\
      \end{array}
    \right)$ is the covariance.

\subsection{Baryon acoustic oscillations}
We employ the BAO points \cite{alam2017clustering} presented in Tab.\ref{tab:1} to constrain the model parameters.
\begin{table}[t]
\begin{center}
  \begin{tabular}{|c|c|c|c|c|c|}
\hline $D_{V}(r_{d,fid}/r_{d})$&$D_{M}(r_{d,fid}/r_{d})$&$H(r_{d}/r_{d,fid})$&$D_{V}(r_{d,fid}/ r_{d})$&$D_{M}(r_{d,fid}/ r_{d})$&$H(r_{d}/r_{d,fid})$
\\
 $z=0.32$ & $z=0.32$& $z=0.32$&$z=0.57$&$z=0.57$&$z=0.57$\\
 $[Mpc]$ &$[Mpc]$&$[km$ $s^{-1}Mpc^{-1}]$&$[Mpc]$&$[Mpc]$&$[km$ $s^{-1}Mpc^{-1}]$\\
\hline
$1270\pm14$&$1294\pm21$&$78.4\pm2.3$&$2033\pm21$&$2179\pm35$&$96.6\pm2.4$\\
\hline
\end{tabular}
  \caption{This is the BOSS DR12 sample, where $D_{V}=(z D_{M}^{2}/H)^{\frac{1}{3}}$ is the dilation scale, $r_{d,fid}=147.78$ Mpc is the fiducial sound horizon, $r_{d}=r_{s}(z_{d})$ is the sound horizon at the drag epoch $z_d$, $z_d$ can be calculated by using $z_{d}=1291\frac{\omega_{(m)}^{0.251}}{1+0.659\omega_{(m)}^{0.828}}(1+b_{1}\omega_{(b)}^{b_{2}})$, $b_{1}=0.313\omega_{(m)}^{-0.419}(1+0.607\omega_{(m)}^{0.674})$, $b_{2}=0.238\omega_{(m)}^{0.223}$ \cite{Eisenstein1997Baryonic}.}
  \label{tab:1}
\end{center}
\end{table}

The BAO likelihood for each parameter combination are
\begin{equation}
  \chi_1^2=\sum_{i=1}^{2}\frac{D_{V}(z_i)(r_{d,fid}/r_{d})-(D_{V}(r_{d,fid}/r_{d}))_{obs}(z_i)}{\sigma_{1i}^2},
\end{equation}

\begin{equation}
  \chi_2^2=\sum_{i=1}^{2}\frac{D_{M}(z_i)(r_{d,fid}/r_{d})-(D_{M}(r_{d,fid}/r_{d}))_{obs}(z_i)}{\sigma_{2i}^2},
\end{equation}

\begin{equation}
  \chi_3^2=\sum_{i=1}^{2}\frac{H(z_i)(r_{d}/r_{d,fid})-(H(r_{d}/r_{d,fid}))_{obs}(z_i)}{\sigma_{3i}^2}.
\end{equation}

Therefore, the total BAO likelihood is
\begin{equation}
  \chi_{BAO}^2=\chi_1^2+\chi_2^2+\chi_3^2.
\end{equation}

\subsection{Type Ia supernovae}
We employ the Joint Light-curve Analysis (JLA) supernova sample \cite{betoule2014improved},\footnote{Although the Pantheon sample \cite{scolnic2018complete} is available now, fitting results won't be too much different if we use the old data.} so the distance modulus is assumed as the following formula
\begin{equation}
  \mu_{obs}=m_B-(M_B-a X_1+b c),
\end{equation}
where $m_B$ and $M_B$ are SN Ia peak apparent magnitude and SN Ia absolute magnitude, respectively. $c$ is the color parameter, and $X_1$ is the stretch factor. Therefore, the likelihood for SN Ia is defined as
\begin{equation}
\chi_{SN}^2=\Delta^{T}C_{SN}^{-1}\Delta,
\end{equation}
where $\Delta=\mu-\mu_{obs}$, and $C_{SN}$ is the covariance matrix.

Finally, we have the following total likelihood
\begin{equation}
   \chi_{tot}^2=\chi_{CMB}^2+\chi_{BAO}^2+\chi_{SN}^2
\end{equation}

\begin{figure}
  \includegraphics[width=1\textwidth]{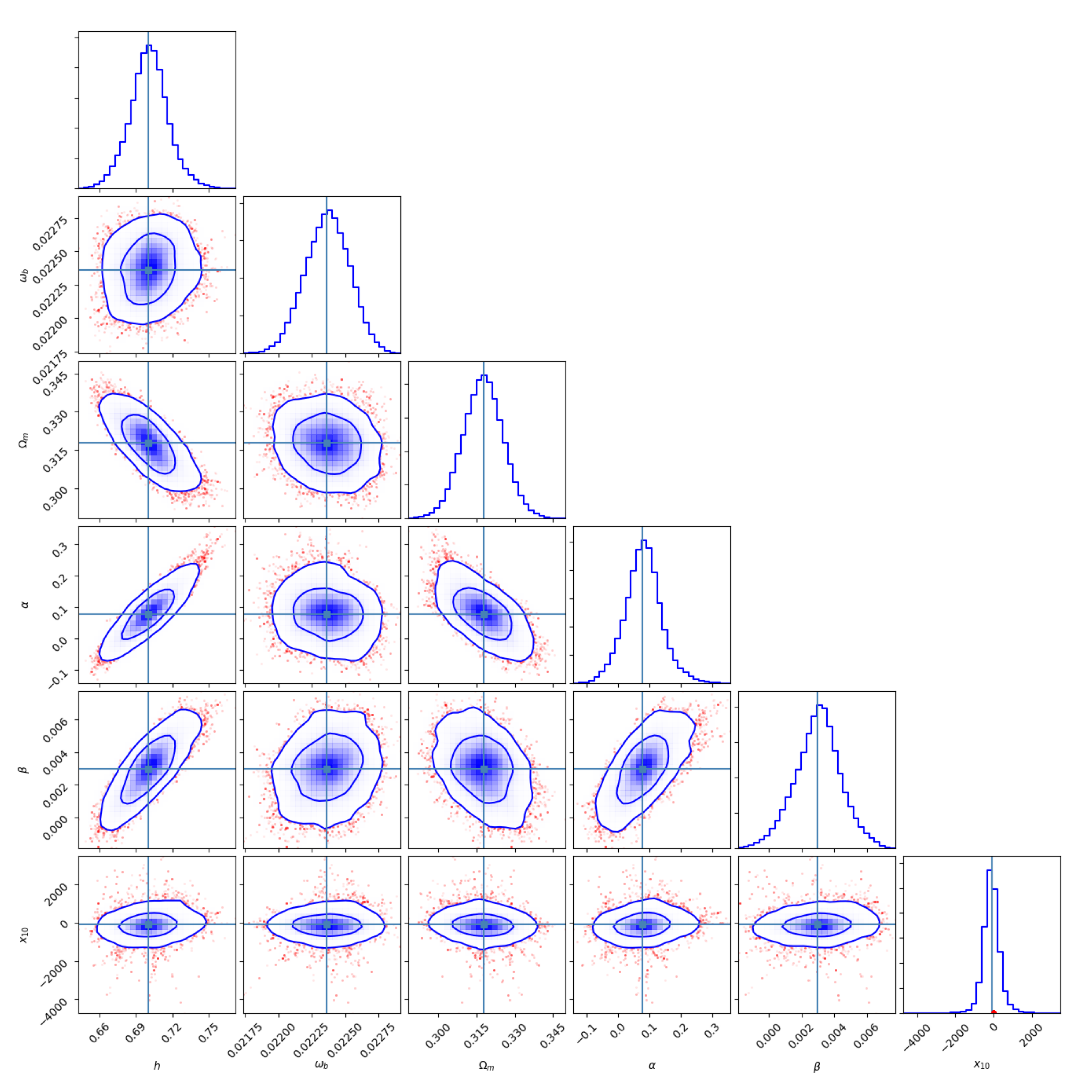}
\caption{$1\sigma$ and $2\sigma$ confidence regions and probability densities for the parameters in the coupled generalized three-form dark energy model.}
\label{fig:1}       % Give a unique label
\end{figure}

\begin{figure}
  \includegraphics[width=0.83\textwidth]{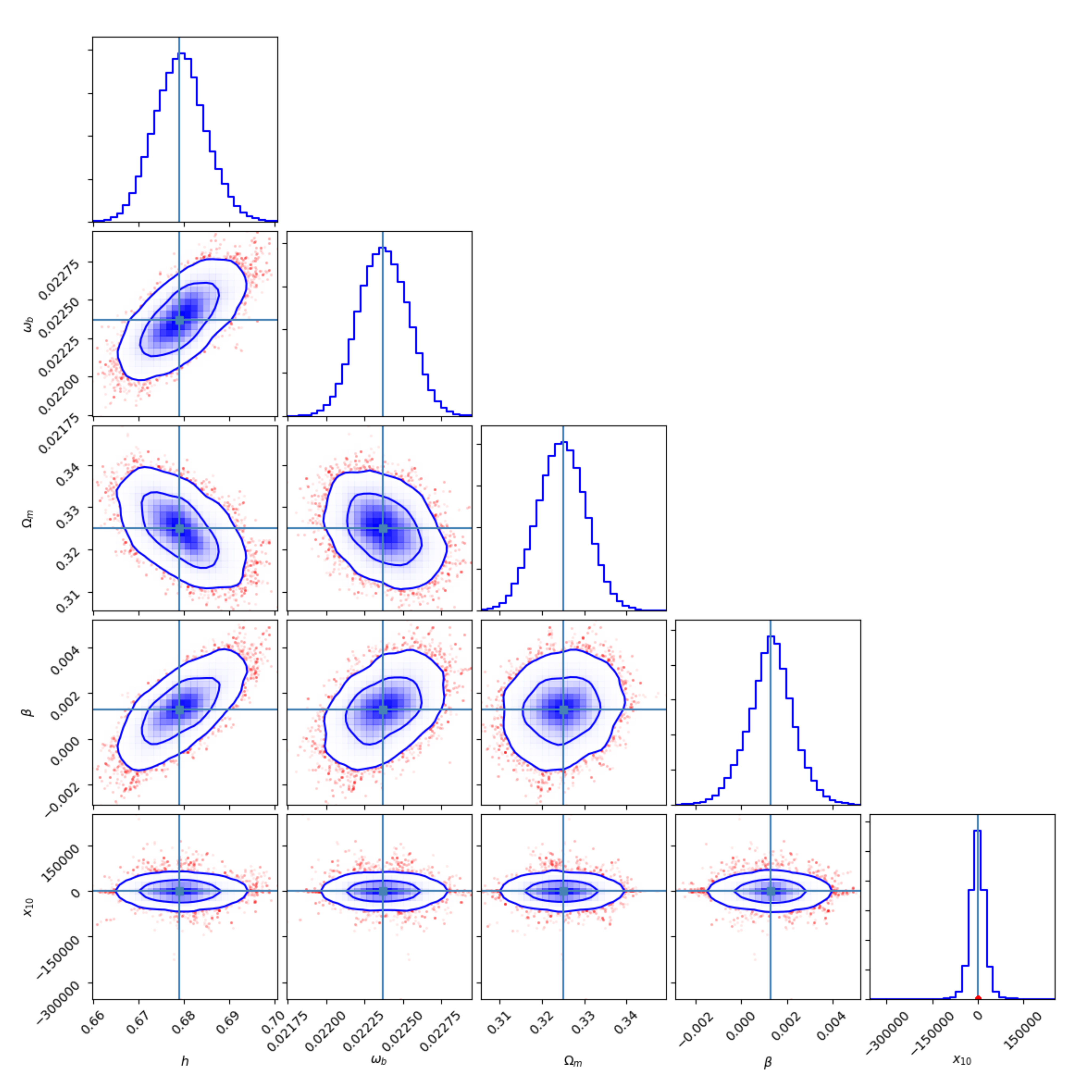}
\caption{$1\sigma$ and $2\sigma$ confidence regions and probability densities for the parameters in the coupled three-form dark energy model.}
\label{fig:2}       % Give a unique label
\end{figure}

\begin{figure}
  \includegraphics[width=0.66\textwidth]{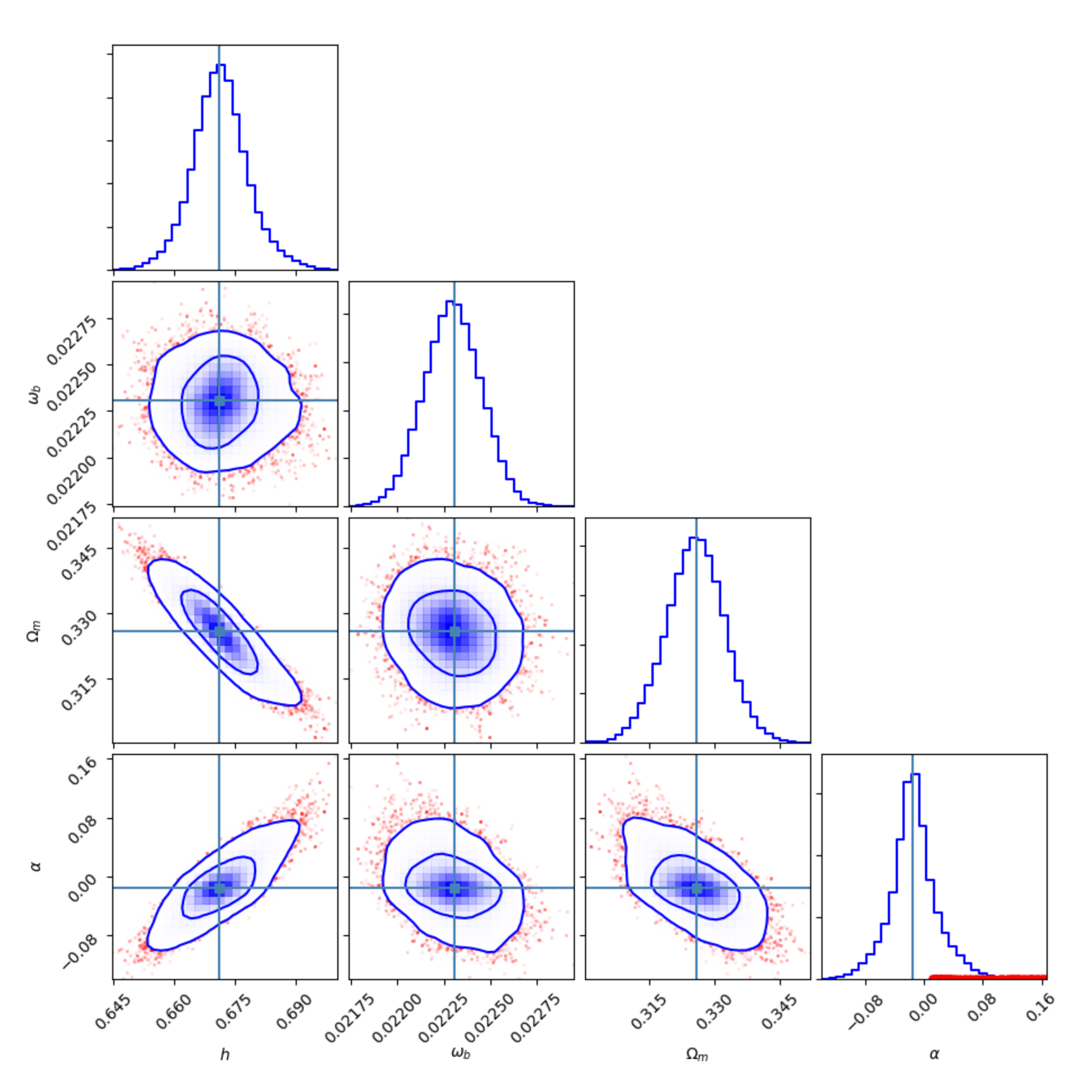}
\caption{$1\sigma$ and $2\sigma$ confidence regions and probability densities for the parameters in the generalized three-form dark energy model.}
\label{fig:3}       % Give a unique label
\end{figure}

\begin{figure}
  \includegraphics[width=0.5\textwidth]{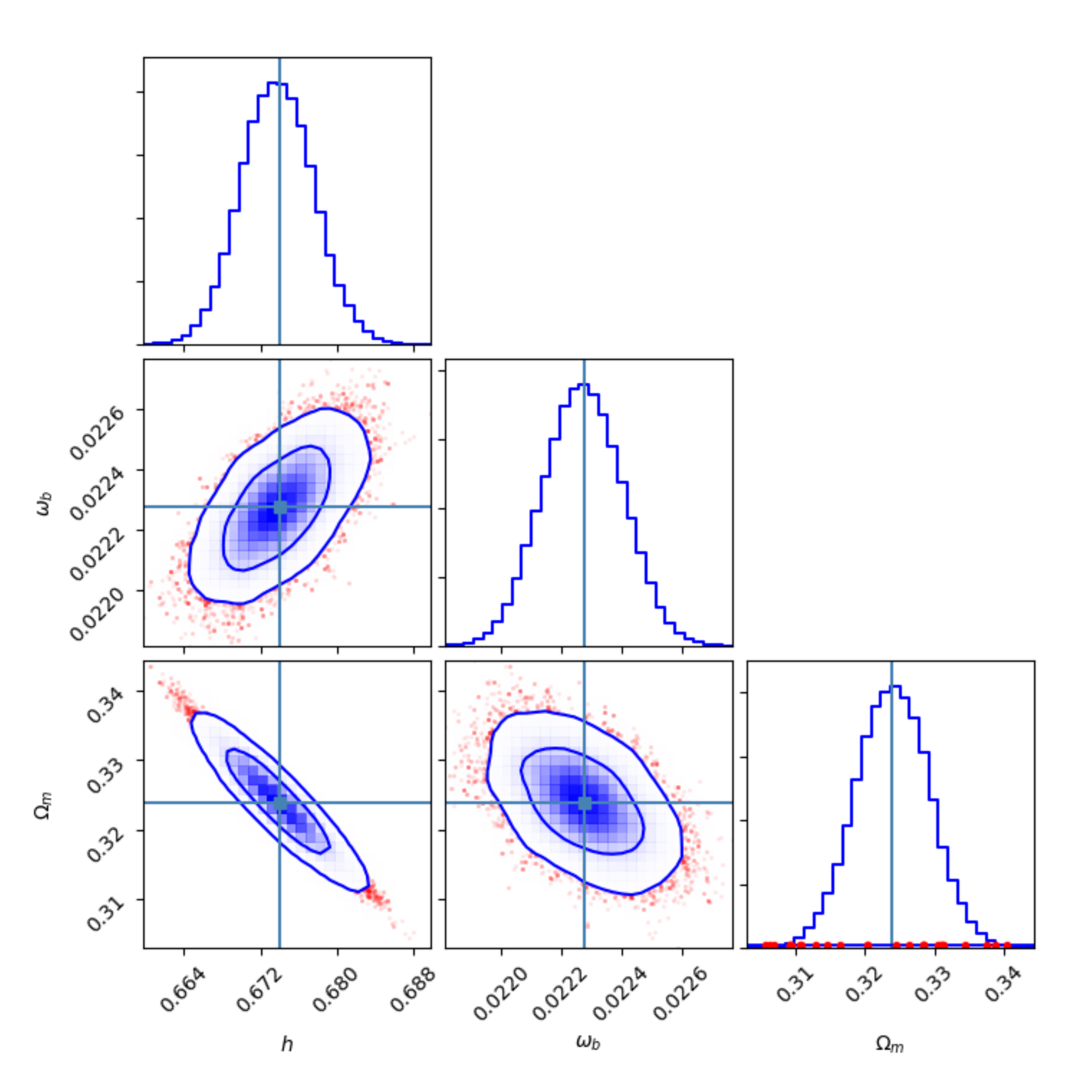}
\caption{$1\sigma$ and $2\sigma$ confidence regions and probability densities for the parameters in the $\Lambda$CDM model.}
\label{fig:4}       % Give a unique label
\end{figure}

In Tab.\ref{tab:2} and Fig.\ref{fig:1}-Fig.\ref{fig:4}, we show the constraint results of cosmological parameters and the tension between the best-fit $H_0$ and R19 in the coupled generalized three-form dark energy model ($\alpha\neq0,\beta\neq0$). For comparison, we also consider the coupled three-form dark energy model ($\alpha=0,\beta\neq0$), generalized three-form dark energy model ($\alpha\neq0,\beta=0$), and $\Lambda$CDM model ($\alpha=0,\beta=0$). We obtain $h=0.701_{-0.015}^{+0.014}$ for the coupled generalized three-form dark energy model,  $h=0.679_{-0.005}^{+0.005}$ for the coupled three-form dark energy model, $h=0.671_{-0.006}^{+0.006}$ for the generalized three-form dark energy model, and  $h=0.674 _{-0.004}^{+0.004}$ for the $\Lambda$CDM model. Corresponding, the $H_0$ tension between them and R19 are 2$\sigma$ for the coupled generalized three-form dark energy model, 4.1$\sigma$ for the coupled three-form dark energy model, and 4.5 $\sigma$ for the generalized three-form dark energy model and the $\Lambda$CDM model. The results indicate that the coupled generalized three-form dark energy model can relieve the $H_0$ tension to some extent, while other three models are still in strong tension with R19. One can be told from Fig.\ref{fig:1}-Fig.\ref{fig:4} that the conclusion is obvious since the parameters $\alpha$ and $\beta$ are both positively correlated with $h$. Although the coupled generalized three-form dark energy model can reduce the
$H_0$ tension from around 4$\sigma$ to 2$\sigma$, the tension still can't be completely resolved. This is not unexpected, in fact, several recent works on coupled dark energy model \cite{di2017can,yang2018interacting,di2020nonminimal,di2020interacting,cheng2020testing,lucca2020tensions,gomez2020update} produce similar results, i.e. $H_0$ gets a bit higher but not enough to completely solve the $H_0$ tension, which is mostly alleviated by a bit larger error bars.

We also consider the Akaike information criterion (AIC) to compare each model. AIC is defined as $\chi_{min}^2 + 2k$, where $k$ denotes the number of cosmological parameters. In fact, we only care about the relative value of AIC
between two different models, i.e., $\Delta$AIC = $\Delta\chi_{min}^2 + 2\Delta k$. A model with a smaller value of AIC
is a more supported model. In this paper, the $\Lambda$CDM model serves as
a reference model. From Tab.\ref{tab:2}, we have $\Delta$AIC=$3.411$ for the coupled generalized three-form dark energy model, $\Delta$AIC=$2.581$ for the coupled three-form dark energy model, and $\Delta$AIC=$1.957$ for the generalized three-form dark energy model. We note that although the coupled generalized three-form dark energy model
is the only one model that can reduce the $H_0$ tension to a more acceptable level, it is least supported by the combined datasets including the Planck 2018 compressed data, the BOSS DR12 sample, and the JLA sample.
 \begin{table}
\begin{center}
\begin{tabular}{|c|c|c|c|c| }
\hline Model & $\alpha \neq0,\beta \neq0$ & $\alpha=0,\beta \neq0$ & $\alpha \neq0,\beta=0$ & $\alpha=0,\beta=0$
\\ \hline
$\Omega_{(m)}$    &$0.318_{-0.008}^{+0.007}$ &  $ 0.325_{-0.005}^{+0.005}$ &  $ 0.326_{-0.007}^{+0.006}$ &$0.324_{-0.005}^{+0.005}$
                     \\
$\alpha$         &  $0.081_{-0.054}^{+0.050}$ & $-$ &  $ -0.015_{-0.024}^{+0.029}$  &$-$
                     \\
$\beta$         &  $0.003_{-0.001}^{+0.001}$ & $0.001_{-0.001}^{+0.001}$ &  $-$   & $-$
                     \\
 $h$          &    $ 0.701_{-0.015}^{+0.014}$ & $0.679_{-0.005}^{+0.005}$ & $ 0.671_{-0.006}^{+0.006}$  & $0.674 _{-0.004}^{+0.004}$
                       \\
 $x_{10}$   &    $-101.315_{-194.760}^{+188.298}$ &  $-739.861_{-5932.031}^{+5164.551}$ & $ -$  & $ -$
                       \\
$\omega_{(b)}$   &    $0.02237_{-0.00017}^{+0.00016}$ &  $0.02237_{-0.00015}^{+0.00016}$ & $ 0.02230_{-0.00015}^{+0.00015}$  & $0.2227 _{-0.00013}^{+0.00013}$
                       \\
\hline
$\chi_{min}^2$ & 684.732  & 685.902 & 687.278 &687.321
  \\
AIC &696.732   & 695.902 & 695.278 &693.321
\\
\hline
$H_0$ tension & 2.0$\sigma$   & 4.1$\sigma$ & 4.5 $\sigma$ & 4.5 $\sigma$
\\
\hline
\end{tabular}
\caption{Fitting results of cosmological parameters and the tension between the best-fit $H_0$ and R19 in the coupled generalized three-form dark energy model, coupled three-form dark energy model, generalized three-form dark energy model, and $\Lambda$CDM model using the CMB+BAO+SN data. }
\label{tab:2}
\end{center}
\end{table}

To investigate the best-fit dynamical behavior of each model, let us rewrite the Hubble parameter as
\begin{equation}
\begin{split}
   H^2 & =\frac{\Omega_{(r)}(1+z)^{4}+\Omega_{(c)}((1+x_{1}^{2})/(1+x_{10}^{2}))^{\frac{\beta}{2}}(1+z)^{3}+\Omega_{(b)}(1+z)^{3}}{1-(
 \alpha^2+x_{1}^2)^{\frac{\alpha}{2}}x_2^2} \\
    & =\Omega_{(r)}(1+z)^{4}+\Omega_{(m)}(1+z)^{3}+(1-\Omega_{(r)}-\Omega_{(m)})\frac{\rho_{(Xeff)}}{\rho_{(Xeff0)}}
\end{split}
\end{equation}
where
\begin{equation}\label{}
\begin{split}
\frac{\rho_{(Xeff)}}{\rho_{(Xeff0)}} &=\frac{1}{1-\Omega_{(m)}-\Omega_{(r)}}(\frac{\Omega_{(r)}(1+z)^{4}+\Omega_{(c)}((1+x_{1}^{2})/(1+x_{10}^{2}))^{\frac{\beta}{2}}(1+z)^{3}+\Omega_{(b)}(1+z)^{3}}{1-(
 \alpha^2+x_{1}^2)^{\frac{\alpha}{2}}x_2^2}\\&-\Omega_{(r)}(1+z)^{4}-\Omega_{(m)}(1+z)^{3})
=\exp\left[\int_{0}^{z}\frac{3(1+\omega_{(Xeff)}(\tilde{z}))}{1+\tilde{z}}d\tilde{z}\right]
\end{split}
\end{equation}
is the normalized effective dark energy density and $\omega_{(Xeff)}$ is the effective equation of state of dark energy.
\begin{figure}
  \includegraphics[width=0.6\textwidth]{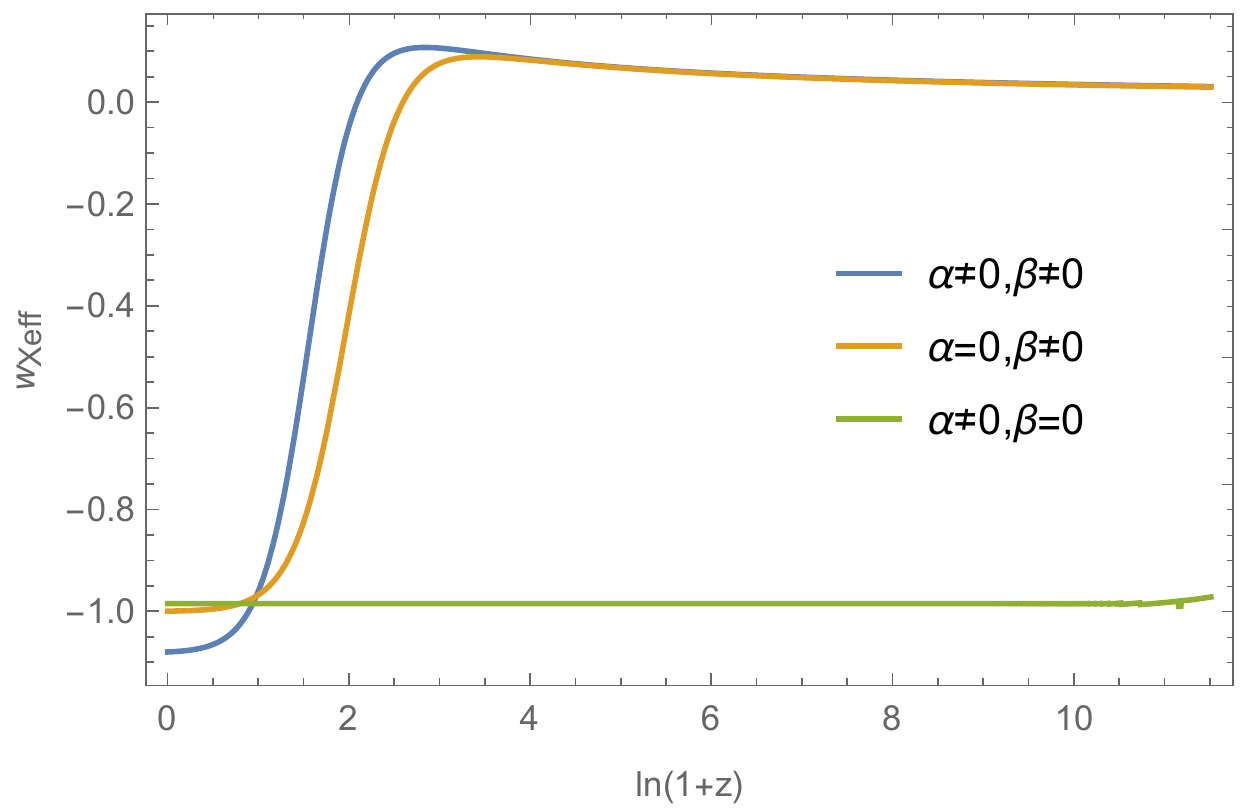}
\caption{The effective equations of state of dark energy in the coupled generalized three-form dark energy model, coupled three-form dark energy model, and generalized three-form dark energy model.}
\label{fig:5}       % Give a unique label
\end{figure}
The effective equations of state of dark energy in the coupled generalized three-form dark energy model, coupled three-form dark energy model, and generalized three-form dark energy model are plotted in Fig.\ref{fig:5}, which shows that the coupled generalized three-form dark energy model is equivalent to a quintom dark energy model, in which dark energy has a small positive equation of state at early epoch, behaving like some form of early dark energy. The Fig.\ref{fig:5} also shows that the effective equation of state of dark energy of the coupled three-form dark energy model have similar behavior with that of the coupled generalized three-form dark energy model at redshift larger than around 30, while the generalized three-form dark energy model can be approximatively regarded as the $w$CDM model with effective equation of state of dark energy slightly larger than $-1$.

\section{Conclusions}
We wish to investigate whether the coupled generalized three-form dark energy model proposed in this paper can resolve the $H_0$ tension. Before applying the datasets, we perform a dynamical analysis on the new model and obtain four fixed points, including a saddle point representing a radiation dominated Universe, a saddle point representing a matter dominated Universe, and two attractors representing two dark energy dominated Universes. We then combine the Planck 2018 compressed CMB data with the BAO data and the JLA data to constrain the model parameters of the coupled generalized three-form dark energy model. For comparison, we also consider the coupled three-form dark energy model, generalized three-form dark energy model, and $\Lambda$CDM model. We find that the coupled generalized three-form dark energy model
is the only one model that can reduce the $H_0$ tension to a more acceptable level, with $H_0=70.1_{-1.5}^{+1.4}$ km/s/Mpc, which is consistent with R19 at $2.0$$\sigma$ confidence level. We also investigate the best-fit dynamical behavior of the coupled generalized three-form dark energy model, and show that the coupled generalized three-form dark energy is equivalent to a quintom, which behaves like some form of early dark energy with a small positive equation of state at early epoch.
\section*{Acknowledgments}
The paper is partially supported by the Natural Science Foundation of China.

\end{document}